\documentclass[a4paper,11pt]{article}
\pdfoutput=1 
\usepackage{jheppub} 
\usepackage[colorlinks=true, linkcolor = blue, urlcolor = blue, citecolor = blue, anchorcolor = blue]{hyperref}
\usepackage{amsmath,amssymb,mathtools,slashed}
\usepackage{multicol,multirow,makecell,caption}
\usepackage{graphicx,xcolor,float,easyReview}
\usepackage[normalem]{ulem}

\title{Revisiting the Inert Scalar Dark Matter with Vector-like Quarks}

\author[a]{Prasanta Kumar Das,}
\author[b]{Shyamashish Dey,}
\author[b]{Saumyen Kundu,}
\author[~b]{Santosh Kumar Rai}

\affiliation[a]{Department of Physics, Birla Institute of Technology and Science-Pilani, K. K. Birla Goa campus, NH-17B, Zuarinagar, Goa 403726, India}
\affiliation[b]{Regional Centre for Accelerator-based Particle Physics, Harish-Chandra Research Institute, A CI of Homi Bhabha National Institute, Chhatnag Road, Jhunsi, Prayagraj 211019, India}

\emailAdd{pdas@goa.bits-pilani.ac.in}
\emailAdd{shyamashishdey@hri.res.in}
\emailAdd{saumyenkundu@hri.res.in}
\emailAdd{skrai@hri.res.in}

\preprint{HRI-RECAPP-2024-07}

 \abstract{The inert doublet model (IDM), a minimal extension of the Standard Model (SM), provides a scalar dark matter (DM) candidate that belongs to the additional Higgs doublet. The model faces challenges in achieving the correct relic abundance for compressed spectra and DM masses in the high-mass range. In this work we introduce a $Z_2$-odd singlet vector-like quark (VLQ) into the IDM framework that helps us alleviate these issues and provide new channels of contributions to the relic abundance. The VLQ not only enhances the DM relic abundance for masses above $~550$ GeV but also eases constraints from direct detection experiments by enabling smaller couplings between the inert scalars and the SM Higgs. We analyze the impact of the VLQ on DM phenomenology, including relic density, direct and indirect detection constraints. The results demonstrate that the extended IDM framework not only resolves existing limitations in the compressed spectrum but also offers exciting prospects for detection in current and future collider experiments.}

\begin{document}
\maketitle
\section{Introduction}\label{sec:intro}

The Standard Model (SM) of Particle Physics remains incomplete in addressing the need for a viable candidate for dark matter (DM), despite the strong observational evidence supporting its existence. Various models have tried to address this gap by proposing potential DM candidates, and its possible signatures in the context of ongoing dark matter search experiments. Notably the DM candidates can be classified based on their mass, coupling or production mechanism in the early Universe. Weakly interacting massive particles (WIMP) belong to one such class that has been widely studied and continues to be a compelling candidate despite being constrained by stringent direct detection (DD) bounds~\cite{PRL.127.261802, PRL.131.041002}. Among the various proposed models is the \emph{inert doublet model} (IDM), which features an extended scalar sector that becomes inert due to a discrete $Z_2$ symmetry~\cite{PhysRevD.18.2574, PhysRevD.74.015007, Honorez_2007}.

The IDM is one of the simplest extensions of the SM for incorporating a dark matter (DM) candidate, where the lightest neutral component of the $Z_2$-odd doublet serves as the DM particle. Beyond its simplicity, the IDM is widely studied for its rich phenomenology, spanning from cosmology to collider physics. It has been examined in the context of strong first-order phase transitions to potentially explain baryon asymmetry in the Universe~\cite{PRD.86.055001}, thereby suggesting a possible connection with symmetric DM~\cite{1110.5334}. The model’s extended scalar sector can lead to an inert phase via multiple phase transitions~\cite{PRD.82.123533} potentially resulting in multi-peaked gravitational waves~\cite{1910.00717} with detectable signatures in direct detection experiments~\cite{2205.06669}. Additionally, the IDM provides interesting collider signatures and has been extensively explored at both current and future colliders~\cite{PRD.76.095011,PRD.81.035003}. Beside the IDM in its minimal form, there also exist various extensions of IDM in the literature. For example, singlet scalar extension in the context of two-component DM~\cite{2108.08061}, extension with singlet Majorana neutrinos for radiative mass generation of SM neutrinos~\cite{Ma:2006km,PRD.96.115012}, vector-like leptons~\cite{Bandyopadhyay:2023joz} and quark~\cite{Ghosh:2024boo} extension to enhance collider signatures with long-lived searches and/or to boost DM production at the LHC has been considered.

Phenomenologically an interesting sector opens up when we consider very small mass-splitting in the IDM spectrum. This so-called compressed scenario has already been studied in Ref.~\cite{1510.08069,1912.08875, 2108.08061} for pure IDM as well as its extensions. Although the compressed scenario is consistent with the Planck-observed relic density bound~\cite{refId0} for dark matter (DM) masses $\lesssim 100$ GeV, it becomes overabundant for DM masses above $\sim 550$ GeV, while the intermediate mass region remains highly underabundant. In an effort to obtain the correct relic abundance in the higher mass region we examine, in this work, the role of VLQ in the inert DM phenomenology which has not been addressed in previous studies.

The VLQs appear in many extensions of the SM such as composite Higgs~\cite{Paolo_Lodone_2008}, little Higgs~\cite{Arkani-Hamed:2002ikv}, extra dimensions~\cite{Moreau:2006np, delAguila:2000kb} and in grand unified theories (GUTs) like $E_6$~\cite{Dutta:2016ach,Das:2017fjf} (for a recent review see ref.~\cite{Alves:2023ufm} and references therein). The VLQ leads to interesting phenomenological implications if charged under an existing discrete symmetry that stabilizes the DM in various models~\cite{Chala:2018qdf,Cornell:2022nky}. We exploit this feature here in stabilizing the DM and utilize its color interactions to obtain correct relic abundance of the inert DM in the higher mass region.

The primary focus of this work is to explore the interesting scenario of a small mass-splitting between the inert scalars and to investigate how an additional $Z_2$-odd singlet VLQ influences scalar DM phenomenology. We show that the VLQ enables us to have smaller values of the coupling between SM Higgs and inert scalars thereby avoiding the DD constraints. We also discuss in detail how effectively the VLQ improves the relic abundance and opens up new IDM parameter region which might lead to interesting signatures at present and future collider experiments. 

In this work we perform a phenomenological analysis of IDM with additional VLQ in a model independent way. As pointed above VLQ can appear in GUT theories. Some notable examples are the irreducible representation {\bf10} of $SO(10)$ which accommodates down-type VLQs when broken to $SU(5)$~\cite{Dorsner:2014wva}. Similarly, both singlet up-type and down-type VLQs arise in the popular $SU(5)_R \times SU(5)_L$ unification scheme  \cite{JHEP02(2017)080}. Regarding the $Z_2$ symmetry of the VLQ and IDM, we note that breaking $SO(10)$ via a Higgs multiplet with an even charge under its $U(1)_X$ subgroup can leave behind a remnant parity symmetry analogous to $Z_2$~\cite{PhysRevD.80.085020}. Additionally, an explicit $Z_2$ symmetry can be imposed at the GUT scale~\cite{Kephart:2015oaa}. Although this discussion does not present a complete theoretical framework, it serves as a useful starting point for exploring theories underlying our model. 

The paper is organized as follows: in section~\ref{sec:model} we describe the model and its parameters and then discuss different theoretical and experimental constraints on the masses and couplings of the model in section~\ref{sec:const}. In section~\ref{sec:dmpheno}, we first inspect the DM phenomenology at direct and indirect detection experiments and the constraints imposed on the model. We then discuss phenomenology from the relic density perspective and the implications of the additional VLQ on the compressed DM scenario. Finally, section~\ref{sec:concl} concludes the discussion. 

\section{Model}\label{sec:model}
The inert doublet model extends the SM by adding an additional scalar doublet, $\eta$, to the existing scalar sector, $\Phi$. The model includes an additional $Z_2$ symmetry, under which the SM scalar doublet is even, while the new doublet is odd. This symmetry prevents the new doublet from coupling to SM matter fields, rendering it `inert'. The scalar potential is given by
\begin{align}\label{eq:pot}
	V(\Phi,\eta) = \mu_1^2 {\Phi}^\dag\Phi + \mu_2^2 {\eta}^\dag\eta + \lambda_1 ({\Phi}^\dag\Phi)^2 
	+ \lambda_2 ({\eta}^\dag\eta)^2 &+ \lambda_3( {\Phi}^\dag\Phi)({\eta}^\dag\eta) +\lambda_4( {\eta}^\dag\Phi)({\Phi}^\dag\eta) \nonumber \\
	&+ \frac{\lambda_5}{2}\left[( {\eta}^\dag\Phi)^2 + h.c.\right] 	
\end{align} 

Here, the parameter $\mu_1^2$ is chosen to be negative which drives the spontaneous symmetry breaking (SSB) by providing vacuum expectation value (VEV) to the neutral CP-even component of $\Phi$. The mass parameter $\mu_2^2$ is positive which ensures that $\eta$ does not obtain any VEV and the $Z_2$ symmetry is unbroken. This essentially ensures the stability of the lightest (pseudo-)scalar making it a suitable DM candidate in the model.  %

We modify this model by including an additional VLQ that is $SU(2)_L$-singlet with electric charge of either $+\frac{2}{3}$ or $-\frac{1}{3}$. The VLQ which is also odd under the $Z_2$ symmetry does not mix with the SM quarks. It interacts with the SM via the inert doublet through a Yukawa interaction given by %
\begin{equation}\label{eq:yukawa}
	\mathcal{L}_{Y} = y_{\xi}~ \overline{Q}_L\eta\,\xi + h.c.
\end{equation}
where $y_\xi$ is the coupling strength. The VLQ has mass $m_\xi$ which is given by the mass term in the Lagrangian, $m_{\xi}\;\overline{\xi}\,\xi$. Although the VLQ can couple to all quark generations, we assume here that it couples only to the third generation of SM quarks via the inert doublet.  This ensures compliance with flavor constraints, which require that the heavy colored field interacts with only one generation of the SM quark family.

\begin{table}[t!]
	\centering
		\begin{tabular}{|>{\centering\arraybackslash} m{3.0cm}|>{\centering\arraybackslash}m{1.25cm}>{\centering\arraybackslash}m{1.25cm}>{\centering\arraybackslash}m{1.25cm}c|}\hline
			
			\textbf{Fields}&  \multicolumn{3}{c}{\boldmath $\mathbf{\underbrace{ SU(3)_C\otimes SU(2)_L \otimes U(1)_{Y}}}$} & {\boldmath $\otimes\;Z_2$} \\ \hline\hline
			\multicolumn{5}{|l|}{Fermions:} \\ \hline
			$Q_L=\left(\begin{matrix} u \\ d \end{matrix}\right)_{L}$ & 3 & 2 & $\phantom{-}\frac{1}{3}$ & $+$\\
			\hline
			$u_R$ & 3 & 1 & $\hphantom{-}\frac{4}{3}$ & $+$ \\[1pt]
			\hline
			$d_R$ & 3  & 1 & $-\frac{2}{3}$ & $+$ \\
			\hline
			$L_L=\left(\begin{matrix} \nu_\ell \\ \ell  \end{matrix}\right)_L$ & 1  & 2 & $-$1 & $+$ \\
			\hline
			$L_R$ & 1  & 1 & $-2$ & $+$ \\
			\hline
			$\xi: (\xi_L,~\xi_R)$ & 3  & 1 & $\phantom{-}\frac{4}{3}$ & $-$ \\
			\hline\hline 
			\multicolumn{5}{|l|}{Scalars:} \\ \hline
			$\Phi = \left(\begin{matrix} \phi^+\\ \phi^0 \end{matrix}\right)$ &  1 & 2 & \phantom{-}1 & $+$ \\
			\hline
			$\eta = \left(\begin{matrix} \eta^+ \\ \eta^0 \end{matrix}\right)$ &  1 & 2 & \phantom{-}1 & $-$ \\
			\hline
		\end{tabular}
		\caption{The fields of the model considered and their quantum number under the SM gauge group and the additional $Z_2$ symmetry. Here $\xi$ is the VLQ and $\eta$ is the new doublet. The definition of electromagnetic charge for all the fields follows the equation $Q_{EM} = T_3 + \frac{Y}{2}$. }
		\label{tab:tab1}
	\end{table}
	
The independent parameters from the IDM are $\{\mu_1^2, \mu_2^2,\lambda_1,\lambda_2, \lambda_3, \lambda_4, \lambda_5\}$ and from the VLQ sector $\{y_\xi, m_\xi\}$. After symmetry breaking, the masses of the physical scalar fields are given as, 
	\begin{align} \label{eq:massh}
		m_h^2 &= -2\mu_1^2 = 2\lambda_1 v^2 \\
		m_{\eta_D^+}^2 &= \mu_2^2 + \frac{1}{2}\lambda_3v^2 \label{eq:massHp}\\
		m_{\eta_D^0}^2 &= \mu_2^2 + \frac{1}{2}(\lambda_3 + \lambda_4 + \lambda_5)v^2 = \mu_2^2 + \lambda_L v^2 \label{eq:massH}\\
		m_{\eta_D^A}^2 &= \mu_2^2 + \frac{1}{2}(\lambda_3 + \lambda_4 - \lambda_5)v^2 = \mu_2^2 + \lambda_{\bar{L}} v^2 \label{eq:massA}
	\end{align}
\noindent 
where, $\lambda_{L,\bar{L}} = \frac{1}{2}(\lambda_3 + \lambda_4 \pm \lambda_5)$, and $h$ is identified as the SM Higgs. The mass relations shown above help in deciding the DM candidate and for our analysis we consider $\lambda_{L} < \lambda_{\bar{L}}$, \textit{i.e.},  $\lambda_5<0$. This choice of the parameters make the CP-even scalar $\eta_D^0$ the lightest particle in the dark sector, making it is our DM candidate, while $\eta_D^A$ and $\eta_D^+$ are the $Z_2$-odd pseudoscalar and charged-scalar, respectively.  Henceforth, the free parameters from the  pure IDM sector will be ($\lambda_2$, $\lambda_L$, $m_{\eta_D^0}$, $m_{\eta_D^A}$, $m_{\eta_D^+}$) while those from the VLQ are its mass ($m_{\xi}$) and the strength of its coupling to the SM and IDM sector ($y_\xi$).

\section{Theoretical and Experimental Constraints}\label{sec:const}
\paragraph{Vacuum Stability --} 
Vacuum stability requires the potential to remain positive at large field values. This is guaranteed by demanding the scalar potential be bounded from below. For the potential in eq.~\refeq{eq:pot}, the individual quartic terms require that 
\begin{equation}\label{eq:vacstab1}
	\lambda_1,\lambda_2 > 0 .
\end{equation}
To prevent the potential from getting unbounded in directions with large values of both $\Phi$ and $\eta$ we need to ensure that 
\begin{equation}\label{eq:vacstab2}
	\lambda_3+2\sqrt{\lambda_1\lambda_2} > 0 ;\nonumber\\
	\lambda_3+\lambda_4-|\lambda_5|+2\sqrt{\lambda_1\lambda_2} > 0 .
\end{equation}
These two equations takes care of the vacuum stability at large values in both the field directions. 
	
\paragraph{Perturbativity --} 
To keep the perturbative calculations within valid limit so that the one-loop corrections are sub-dominant to the tree-level contributions~\cite{PRD.80.123507}, the quartic scalar couplings need to be small enough as given by 
\begin{equation} |\lambda_i|\lesssim 4\pi,\quad i=1,2,3,4,5 . \end{equation}
Also, the Yukawa interaction between the VLQ and the inert doublet is also ensured to be perturbative by the condition
\begin{equation} |\lambda_\xi|\lesssim \sqrt{4\pi}. \end{equation}
	
\paragraph{Electroweak precision observables --} 
Additional scalars and fermions with gauge couplings can alter the electroweak precision observables (EWPO). The $Z_2$-odd VLQ does not mix with the SM quarks and therefore does not affect the EWPO~\cite{PRD.47.2046}. Whereas, the inert doublet, $\eta$, can produce deviations on the oblique parameters ($S$ and $T$) from the SM predictions~\cite{PRD.83.055017}. While the correction on $S$ is negligibly small, the one in $T$ is given by 
\begin{equation}\label{eq:Tpara} 
	\Delta T = \frac{g_2^2}{64\pi^2m_{W}^2}\left[\zeta\left(m_{\eta_D^\pm}^2,m_{\eta_D^A}^2\right) + \zeta\left(m_{\eta_D^\pm}^2,m_{\eta_D^0}^2\right) + \zeta\left(m_{\eta_D^A}^2,m_{\eta_D^0}^2\right) \right]
\end{equation}
with 
\begin{equation}
	\zeta\left(x, y\right)= \begin{cases} \frac{x+y}{2} - \frac{xy}{x-y} \ln\left(\frac{x}{y}\right), & \text{if } x\neq y,\\
		0,              & \text{if } x = y.\end{cases}
\end{equation}
	
The $\Delta T$ depends on the mass differences in the scalar spectrum and therefore puts strong limits on any large mass-splittings between them. The current bound on the oblique electroweak $T$-parameter is 
\begin{equation} \Delta T = 0.07\pm0.12.\end{equation}

In this work we focus on a very compressed mass spectrum for the inert doublet scalars and are not affected by the constraints from the EWPO. 

\paragraph{Flavor Constraints --}

Exotic quarks can contribute to the $B_q^0-\overline{B}_q^0$ mixing via the Cabibbo-Kobayashi-Maskawa (CKM) matrix and modifying the width and mass difference between the heavy (H) and light (L) states. The current bound on these observables are~\cite{HFLAV:2022esi, Albrecht:2024oyn} 
\begin{align}
	&\Delta\Gamma_d/\Gamma_d = 0.001\pm0.010 &\Delta m_d = 0.5065\pm0.0019 \text{ ps}^{-1} \\
	&\Delta\Gamma_s/\Gamma_s = 0.128\pm0.007 &\Delta m_s = 17.765\pm0.004 \text{ ps}^{-1}
\end{align}

However in our case the VLQ being $Z_2$-odd does not mix with the SM quarks, therefore does not modify the CKM parameters. Therefore, as discussed in Sec~\ref{sec:model}, since it only couples to third generation of SM quarks, it does not generate additional box diagrams via charged current interactions contributing to the $B_q^0-\overline{B}_q^0$ oscillation at 1-loop.

The charged-scalars of the inert doublet does not participate at 1-loop corrections of the $b\to s\gamma$ branching ratio due to its odd $Z_2$ charge making the model safe from flavor constraints.

\paragraph{Collider bounds --} 
	
As the IDM is a $SU(2)$ doublet, it couples with the SM gauge bosons with EW coupling strength. 
This puts a strong constraint on the mass of the IDM scalars when they are lighter than the SM gauge bosons.  The precise width measurements of the SM vector bosons require that the $W^\pm$ and $Z$ do not decay to the new scalars. 

Unless a nearly degenerate scenario, this gives~\cite{PRD.76.095011}   
\begin{equation}\label{eq:Zdecay} m_{\eta_D^0} + m_{\eta_D^A} > m_Z. \end{equation}
The mass of the charged-scalar, $m_{\eta_D^\pm}$, is constrained from LEP data below 80 GeV \cite{ParticleDataGroup:2024cfk}, i.e.,
	
\begin{equation}\label{eq:Wdecay} 
	m_{\eta_D^\pm} > 80 \text{ GeV}. 
\end{equation}	

The SM Higgs invisible decay to the inert scalars is constrained to be less than 10\% \cite{PRD.105.092007,ATLAS:2022tnm}.
To avoid this constraint we restrict ourselves to DM masses satisfying 
\begin{equation}\label{eq:hinv} m_{\eta_D^0} > m_h/2 \end{equation}
Despite the DM mass satisfying eq.~\refeq{eq:hinv}, the additional scalar doublet can still influence the $h\to\gamma\gamma$ width via $\eta_D^\pm$ at one-loop depending on $\lambda_3$~\cite{PhysRevD.76.095011,PhysRevD.88.035019,PhysRevD.85.095021}. This puts a modest constraint on $|\lambda_3| \lesssim 1$ for $m_{\eta_D^\pm}>m_Z/2$. Since, we are working in the region with DM mass $\gtrsim550$ GeV, the constraints in eqs.~\refeq{eq:Zdecay}-\refeq{eq:hinv} are automatically satisfied.
	
The LHC will produce the VLQ with large cross-section through strong interaction. The conventional bounds on VLQ depend on its weak decays to a SM quark and gauge bosons, that then further decay semileptonically to give charged leptons and jets in the final state. However, these bounds on the VLQ mass from the conventional channels are not valid for $\xi$ in our model, due to their $Z_2$-odd nature. The VLQ's in our case would decay to a SM quark and the inert scalars, giving a multi-jet final state with large missing transverse energy (MET). Therefore, effective bounds on the VLQ can be interpreted from searches that have been carried out for supersymmetric particles, namely the squarks that decay into SM quarks and neutralino. For a top-like (bottom-like) VLQ decaying in to a top (bottom) quark and DM, these bounds have been recast in our model~\cite{ATLAS:2024lda} giving
\begin{equation}
	m_\xi > 550\;(650) \text{ GeV}.
\end{equation}
	
\section{Inert Scalar Dark Matter and Vector-Like Quarks}\label{sec:dmpheno}

To study the phenomenology of the scalar DM of the IDM in presence of VLQ's  we implement the model in \texttt{SARAH}~\cite{Staub:2013tta} to generate relevant model files for the analysis. 
The CHO library files generated via \texttt{SARAH} is used in \texttt{MicrOMEGAs}~\cite{Alguero:2023zol} to obtain DM-related observables such as direct detection (DD) cross-sections and relic density. The mass spectrum for the particles have been generated using \texttt{SPheno}~\cite{POROD20122458}.
	
\subsection{Direct Detection}
\begin{figure}[t]
	\centering
	\includegraphics[width=0.495\linewidth]{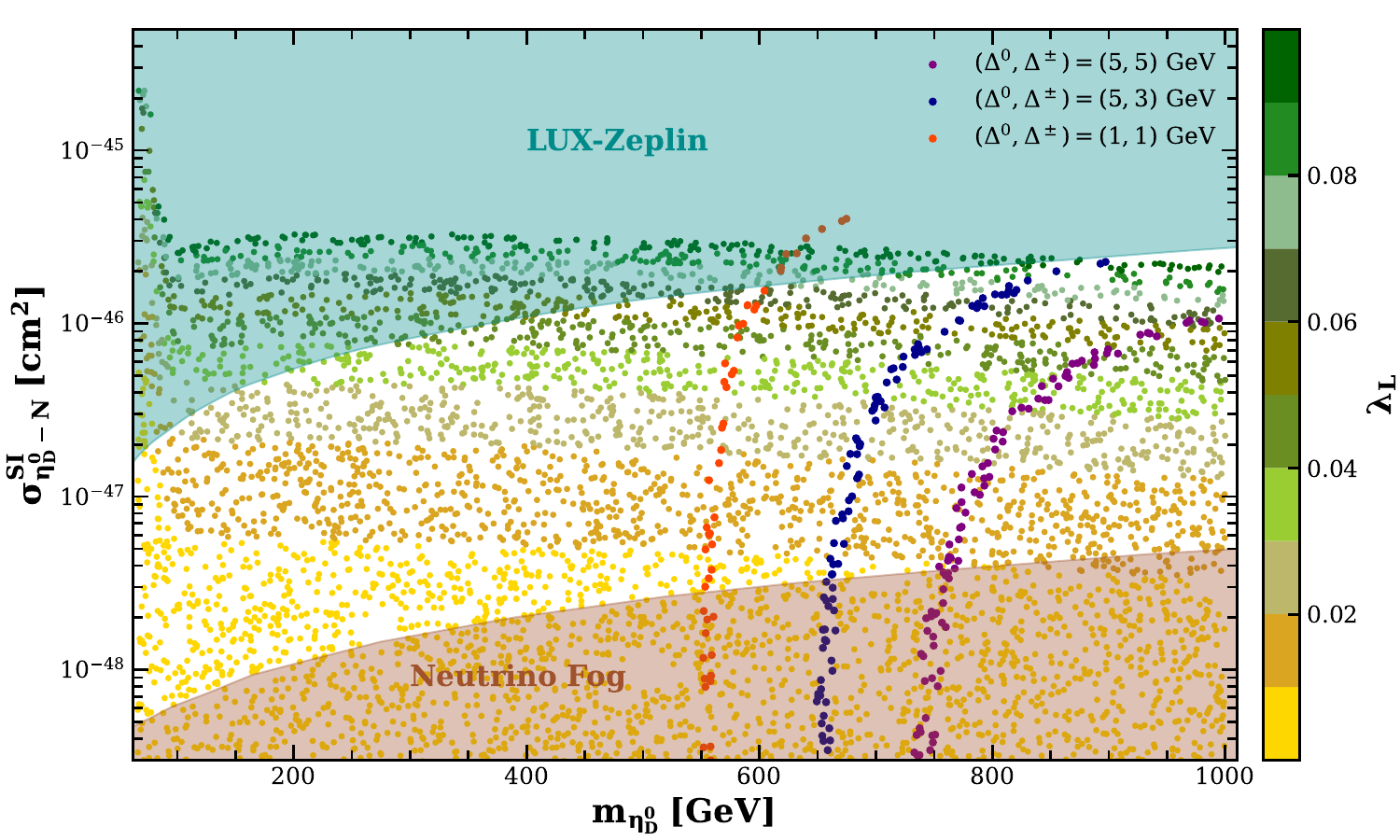}
	\includegraphics[width=0.495\linewidth]{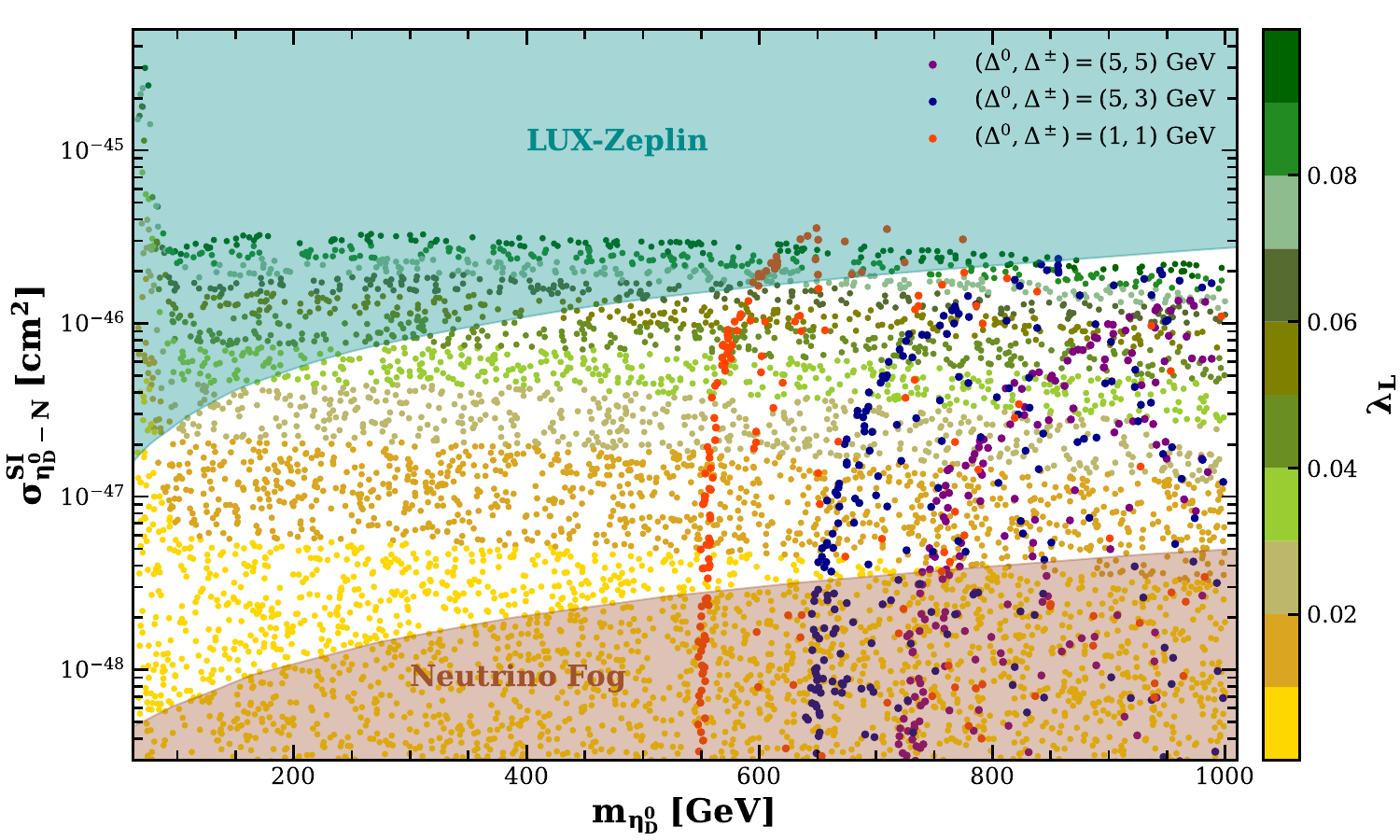}	
	\caption{Spin-independent DM-Nucleon cross-section as a function of DM mass for different randomized value of $\lambda_L$ given by the colorbar for $(\Delta^0,\Delta^\pm)=(5,3)$ GeV. The two shaded regions at the top and bottom of the plot represent the constraints from LUX-Zeplin~\cite{PRL.131.041002} and Neutrino fog~\cite{PRL.127.251802}, respectively. The patches of purple, dark blue and sky blue colored points represent the region with $\Omega h^2 \in [0.118,0.122]$. }
	\label{fig:DD}
\end{figure}
	
Among the dedicated DM search facilities, the direct detection experiments put the most stringent constraints on the WIMP parameter space. In the IDM, the DM interacts with the nuclei of the target materials through its interaction with SM Higgs. In fig.~\ref{fig:DD}, we show the spin-independent (SI) DM-nucleon cross-section at one-loop\footnote{considering contributions from the box diagrams as implemented in MicrOMEGAs~\cite{Alguero:2023zol, Hisano:2015bma}}, $\sigma^{\rm SI}_{\eta_D^0\text{-}N}=\sigma^0_{\eta_D^0\text{-}N}\times f_\Omega$, scaled by the local relic density, $f_\Omega=(\Omega h^2)_{\eta^0_D}/(\Omega h^2)_{\rm Planck}$, as a function of DM mass. Here, $\sigma^0_{\eta_D^0\text{-}N}$ is the cross-section when the DM produces the observed relic abundance. The colormap axis represents the randomly varying $\lambda_L$ values which measures the Higgs-inert scalar interaction. The shaded region at the top is excluded by the LUX-ZEPLIN experiment~\cite{PRL.131.041002}. For a mass-splitting below $\sim100$ keV between the neutral inert scalars for DM mass $m_{\eta_D^0}>m_Z/2$, the inelastic SI cross-section with nuclei shoots up into the DD excluded region~\cite{PhysRevD.74.015007}. We have kept the mass-splitting safely above 100 keV throughout this paper. As we can see in fig.~\ref{fig:DD}, points with $\lambda_L\lesssim 0.01$ is constrained for low-mass DM, whereas for higher value of DM mass the constraint on $\lambda_L$ is comparatively relaxed; for example, $\lambda_L\lesssim0.06$ when the DM mass is $m_{\eta_D^0}\sim 500$ GeV. 
The patches of purple, dark blue and orange colored points have the correct relic density for three different mass-splittings among the dark scalars, denoted by $\Delta^0=m_{\eta_D^A} -m_{\eta_D^0},\text{ and }\Delta^\pm=m_{\eta_D^\pm} -m_{\eta_D^0}$. It is observed that with smaller mass-splitting we require higher value of $\lambda_L$ which for a fixed DM mass pushes the cross-section into the excluded region.  

\begin{figure}[t]
	\centering
	\includegraphics[width=5.5cm]{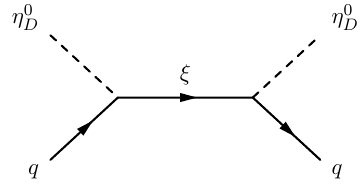} \hspace{0.5cm}
	\includegraphics[width=5.5cm]{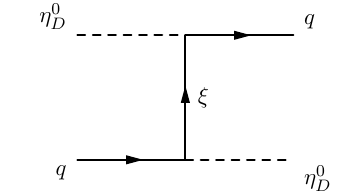} 
	\caption{Tree-level Feynman diagrams contributing to the DM-nucleon scattering mediated by the VLQ.  }
	\label{fig:feynDD}
\end{figure}

The addition of the VLQ, $\xi$, introduce additional diagrams (in figure \ref{fig:feynDD}) by mediating the DM-quark interactions. The scaled SI cross-sections, $\sigma^{\rm SI}_{\eta_D^0\text{-}N}$, for this scenario is depicted in the right panel of figure~\ref{fig:DD} where we see that the constraint on $\lambda_L$  is respected satisfying the observed relic density simultaneously.  
	
It is also worth mentioning that arbitrarily small value of $\lambda_L$ makes the cross-section fall in the neutrino fog region where it becomes difficult to distinguish the DM-nucleon interaction from the neutrino-nucleon ones. However, efforts are being made to devise new techniques to identify a neutrino-mimicking DM signal~\cite{PRL.127.251802}. Therefore, we will  assume that this region can be probed and will choose $\lambda_L$ arbitrarily small or zero being consistent with DD exclusions.

\subsection{Indirect Detection}
	
Indirect detection of dark matter is based on the production of SM particles through DM annihilation or decay, which are detected as excesses in the fluxes of gamma rays, neutrinos, positrons, electrons, and antiprotons. Photons are in particular advantageous since their propagation is unaffected by the interstellar and intergalactic media and thus preserves the spectral and spatial information. Depending on the type of SM particles detected, there are various facilities such as Fermi-LAT~\cite{PRL.115.231301}, MAGIC~\cite{MAGIC_2016}, and H.E.S.S.~\cite{PRL.129.111101} that measure the spectral properties of fluxes from the galactic center and dwarf spheroidal galaxies within their respective operating energy ranges. 
	
\begin{figure}[t]
	\centering
	\includegraphics[width=0.7\linewidth]{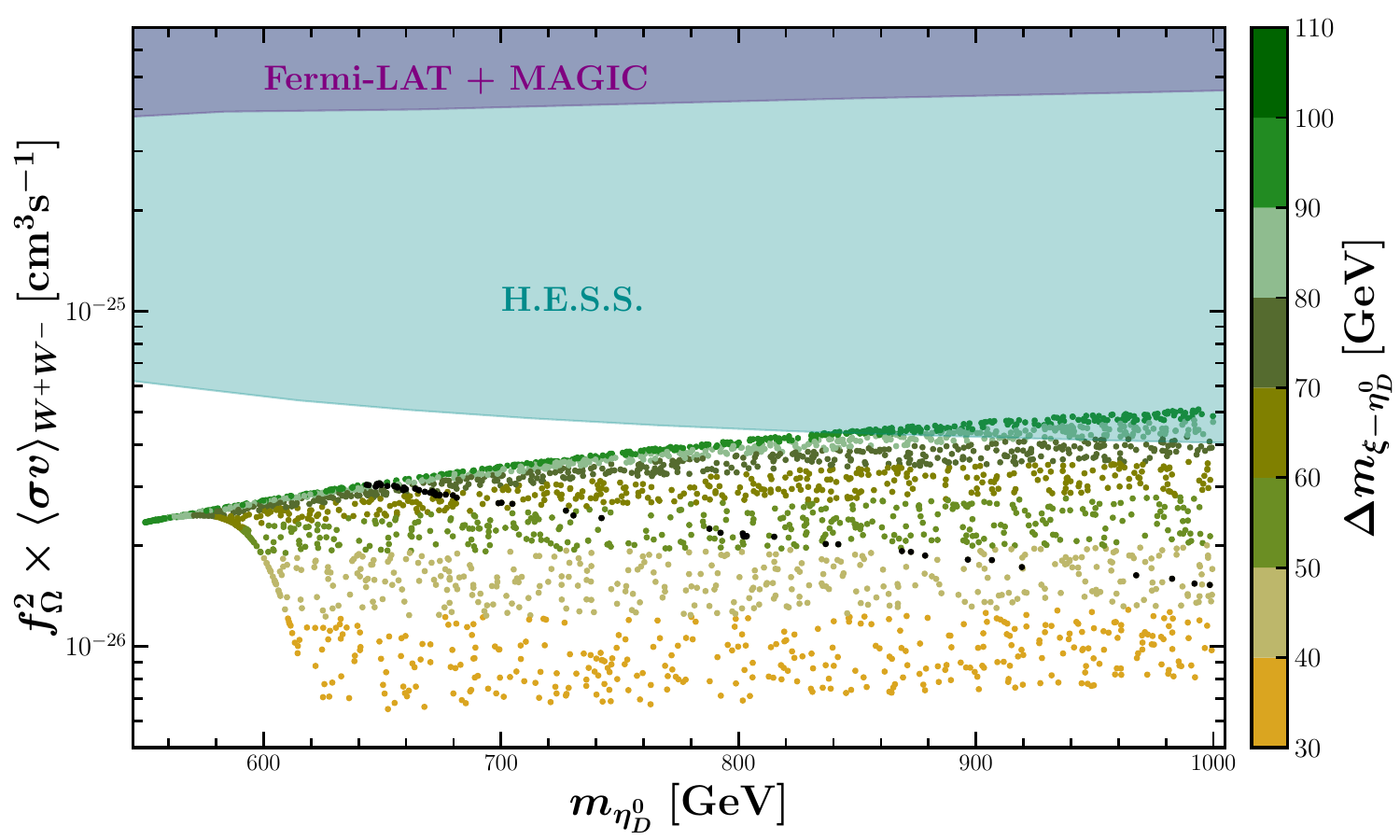}
	\caption{Thermal averaged velocity times cross-section (scaled with $f_\Omega^2$) as a function of DM mass for a random mass gap between the DM and VLQ with $y_{\xi}=0.3$. Here $\lambda_{L}=10^{-3}$ and $\Delta^0=5$ and $\Delta^\pm=3$. The cyan and purple shaded regions are excluded by the results from H.E.S.S.~\cite{PRL.129.111101} and Fermi-LAT and MAGIC~\cite{MAGIC_2016}, respectively. }
	\label{fig:ID}
\end{figure}

For pure IDM the main contribution comes from the four-point gauge interaction into $VV$ with $V\equiv W^\pm,Z$, that eventually result into positron or neutrino or even photon final state. If coming from a source of higher DM density, the flux of these particles can be analyzed to obtain constraints on the model parameter space. Note that the VLQ also introduces additional diagrams, mediating via $t$-channel. The notable contribution from our top-like VLQ would be to the $t\bar{t}$ mode that decays to $b$ quarks and $W$ bosons. However, these contributions will depend on the VLQ mass and the $y_\xi$ coupling strength. In figure~\ref{fig:ID} we show the variation of relic-allowed $\langle\sigma v\rangle$ scaled by $f_\Omega^2$ as a function of DM mass for different values of DM-VLQ mass-gap. We note that the strongest constraint comes from H.E.S.S. via the $W^+W^-$ channel. However, overall indirect detection results do not constrain any additional part of the parameter space in our DM mass region of interest, \textit{i.e.}, $m_{\eta_D^0}\in [500,1000]$ GeV. 
	
\subsection{Relic Density}
	
The WIMP particles that were in thermal equilibrium with the primordial plasma, got decoupled as the Universe expanded and cooled, obtaining a `freeze-out' number density which we know as the relic density. This relic density of the WIMP is sensitive to its annihilation cross-section, mass, and interaction strengths and these properties help us understand and calculate its abundance at current times and also provide constraints on their values in any new physics model.

To find the DM relic density one considers all processes in which the total number of dark sector particles change. Then the DM relic density can be calculated by solving the Boltzmann equation which gives an expression for the changing number density of any given particle over time,
\begin{equation}\label{boltzmn}
	\frac{dn_{\rm DM}}{dt} + 3Hn_{\rm DM} = -\langle\sigma v\rangle_{\rm eff}\Big(n_{\rm DM}^2 - n^2_{\rm eq}\Big),
\end{equation}
where $n_{_{\rm DM}}$ denotes the number density of DM and $n_{\rm eq}= g_{_{\rm DM}}(\frac{m_{_{\rm DM }} T}{2\pi})^{3/2}\exp(-m_{_{\rm DM}}/T)$ is the equilibrium density. For pure IDM case the $\langle\sigma v\rangle_{eff}$ is given by ~\cite{PRD.43.3191,PRD.56.1879}
\begin{eqnarray} 
	\label{sigveffIDM}
	\langle\sigma v\rangle_{eff}^{\rm IDM}&=& \frac{g^2_0}{g^2_{\rm eff}}\langle\sigma v\rangle_{\eta_D^0 \eta_D^0}^{\rm IDM} + \frac{2 g_0 g_i }{g^2_{\rm eff}}\langle\sigma v\rangle_{\eta_D^0 \eta_D^i}^{\rm IDM} \Big(1+\Delta_{\eta_D^i}\Big)^{\frac{3}{2}}  e^{-x \Delta_{ \eta_D^i}} \nonumber \\
	&& + \frac{2 g_i g_j}{g^2_{\rm eff}}\langle\sigma v\rangle_{\eta_D^i\eta_D^j}^{\rm IDM} \Big(1+\Delta_{\eta_D^i}\Big)^{\frac{3}{2}}\Big(1+\Delta_{\eta_D^j}\Big)^{\frac{3}{2}} e^{-x (\Delta_{\eta_D^i}+\Delta_{\eta_D^j})} \hspace{0.5cm}
\end{eqnarray}
where, $g_{\rm eff}=\sum_{i=1}^Ng_i(1+\Delta_i)^{3/2}\exp(-x\Delta_i)$, $\Delta_{i} = \frac{(m_i-m_{\eta_D^0})}{m_{\eta_D^0}}$ and $g_i$ is the multiplicity of the $i^{th}$ particle. As labeled using the particle name as subscripts, the respective contributions in the above equation are from DM-DM annihilation, coannihilation of DM with the remaining $Z_2$ odd scalars and finally the annihilations and coannihilations between the remaining $Z_2$ odd scalars respectively. The exponential term in the last two terms in the above expression represents the suppression when their is a mass-splitting among the coannihhilating particles. So these terms drop off as the mass-splitting increases. Since we focus on small mass-splittings i.e. a compressed spectra, all the terms will contribute to the relic density calculation. 
		
\begin{figure}[t]
	\centering
	\includegraphics[width=0.7\linewidth]{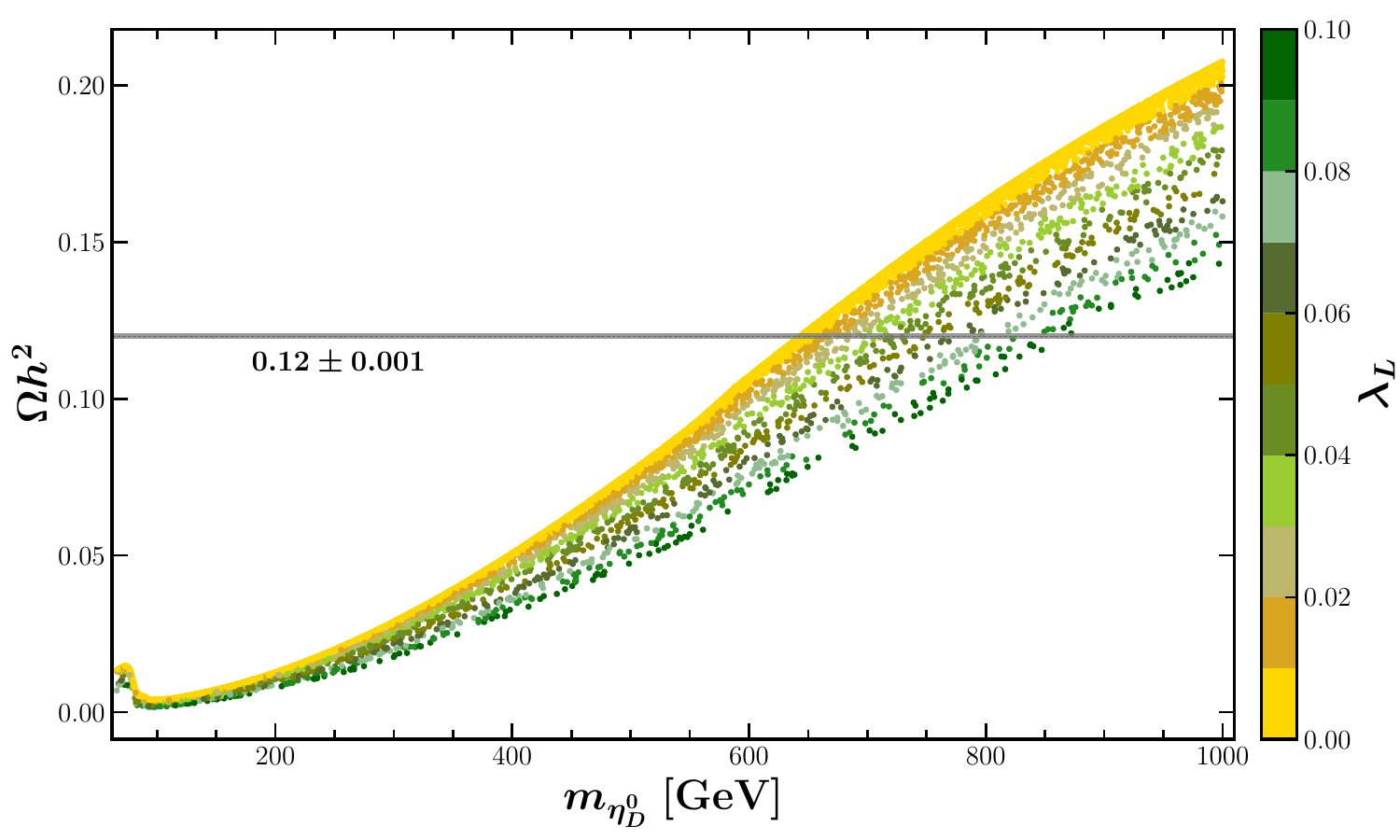}
	\caption{DM relic density as a function of DM mass for different randomized value of $\lambda_L$ given by the colorbar. The black dashed line denotes the Planck-observed value DM relic density.}
	\label{fig:relicfull}
\end{figure}

\begin{figure}[t!]
	\centering
	\includegraphics[angle=0,width=0.48\textwidth]{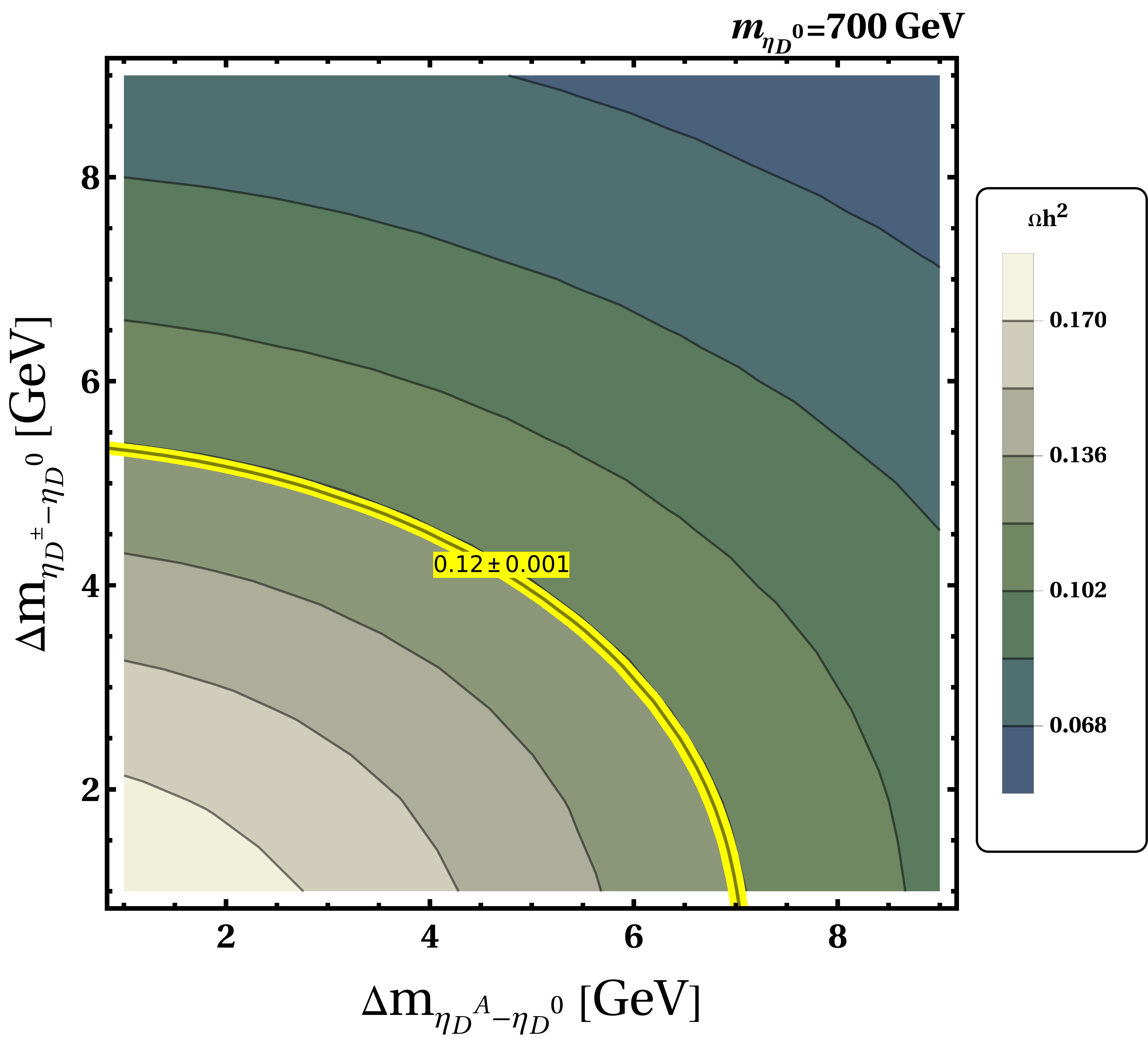} \hspace{.3cm}
	\includegraphics[angle=0,width=0.48\textwidth]{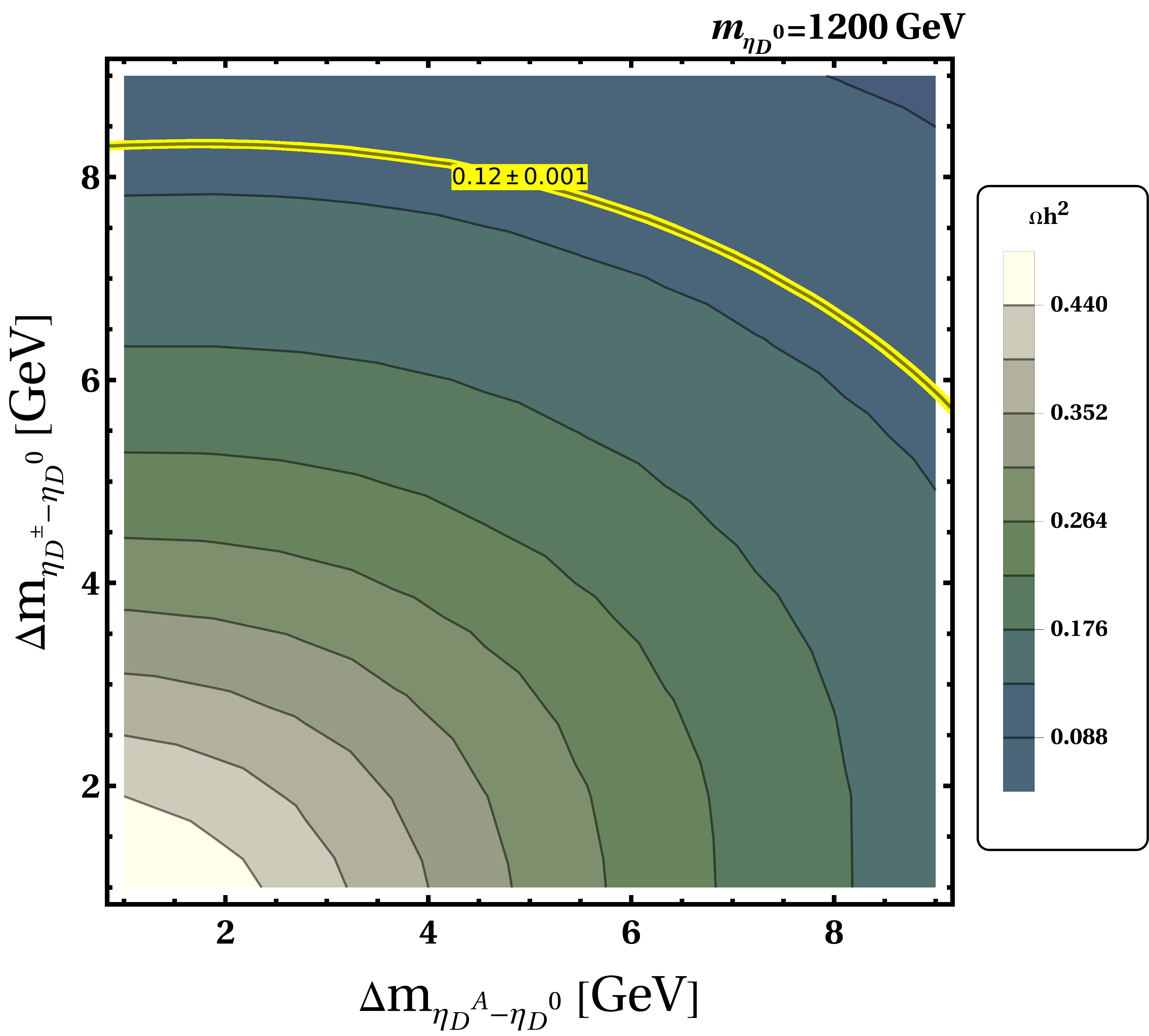}
	\caption{Contour plot of DM relic density for pure IDM case in the plane of $\Delta^0$ and $\Delta^\pm$, for DM masses of 700 GeV (\emph{left}) and 1200 GeV (\emph{right}).}
	\label{fig:contIDM}
\end{figure}

The phenomenology of DM within the framework of the IDM has been extensively studied, including scenarios involving a compressed mass spectrum. We briefly discuss here the implications of a compressed scenario particularly in the high mass region. The annihilation channels as represented by the first term in eq.~\refeq{sigveffIDM} are dominated by $s$-channel Higgs mediated $WW^{(*)}$, $ZZ^{(*)}$ and $b\Bar{b}$ production for $m_{\eta_D^0} < m_W$. As $m_{\eta_D^0}\gtrsim m_W \text{ or } m_Z$, the annihilation channels through the four-point contact vertex of $\eta_D^0-\eta_D^0-V-V$ where $V=W,Z$, open up and dominate the contributions in $\langle\sigma v\rangle_{eff}^{\rm IDM}$, and this continues to be the case even in the mass range above $600$ GeV. However with small mass-splitting, additional contributions to the $\langle\sigma v\rangle_{eff}^{\rm IDM}$ grows as given by the second and third term in eq.~\refeq{sigveffIDM}. While these processes are controlled by mainly gauge-couplings, the mass-splitting plays a crucial role as discussed earlier. In addition, this also enhance the contribution coming from the destructive interference between the four-point contact vertex diagram and the pseudocscalar and charged-scalar mediated annihilations diagrams of DM to the weak gauge boson pair~\cite{Honorez_2007}. It can be seen from fig.~\ref{fig:contIDM} that as we make the mass-splitting in inert sector smaller, the relic abundance increases and eventually crosses the Planck-observed value. Note that even though the smaller mass-splittings will increase the contributions from the second and third term in eq.~\refeq{sigveffIDM} that contribute to decreasing the relic, these are still smaller and the destructive interference wins over. This interplay between the suppression in $\langle\sigma v\rangle_{eff}^{\rm IDM}$ due to the destructive interference and the positive contribution to $\langle\sigma v\rangle_{eff}^{\rm IDM}$ through the second and third terms can be seen from the narrowing of the bands in the fig.~\ref{fig:contIDM} as we go to smaller mass-splittings.   From eqs.~\refeq{eq:massHp},\refeq{eq:massH},\refeq{eq:massA}, the mass-splittings can be evaluated as 
\begin{equation}
	\Delta^0 = m_{\eta_D^A} -m_{\eta_D^0} \approx -\frac{\lambda_5v^2}{2m_{\eta_D^0}}, \hspace{2cm}
	\Delta^\pm = m_{\eta_D^\pm} - m_{\eta_D^0} \approx -\frac{(\lambda_4+\lambda_5)v^2}{4m_{\eta_D^0}} \,\,\, .
\end{equation}
While $\Delta^0$ can be made arbitrarily small by imposing an additional approximate global $U(1)$ symmetry on the inert doublet, arbitrarily small $\Delta^\pm$ requires the imposition of a global $SU(2)$ symmetry~\cite{1510.08069}. However, one notes that even with the additional $SU(2)$ symmetry imposed or extending the SM custodial symmetry to $\eta$, arbitrarily small mass-splittings ($\Delta^\pm$) are difficult to maintain. This is due to the $g^2g^{\prime2}$ term present in the 
one-loop renormalization group equation (RGE) of $\lambda_4$ that breaks the additional symmetries explicitly. $\Delta^\pm$ would also get a non-zero contribution from electroweak symmetry breaking at one-loop. In this work we choose $\Delta^0 > 100$ keV that help us evade the DD constraints and we have kept $\Delta^\pm \gtrsim 1$ GeV. 

The DM relic density for a pure IDM case is found to increase with DM mass. As the annihilation cross-section starts to fall further, the DM relic eventually becomes over-abundant at sufficiently higher mass. This remains the case unless we further increase the mass gap between the IDM particles to suppress the destructive interference or increase the coupling ($\lambda_L$) of DM with the SM Higgs. However, $\lambda_L$ cannot be increased arbitrarily as DD bounds restrict it. Therefore, in the compressed scenario the higher mass region remains over-abundant. 
	
\begin{figure}[t!]
	\centering
	\includegraphics[width=4.9cm]{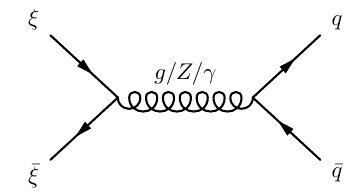} 
	\includegraphics[width=4.9cm]{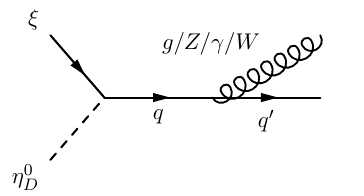} 
	\includegraphics[width=4.9cm]{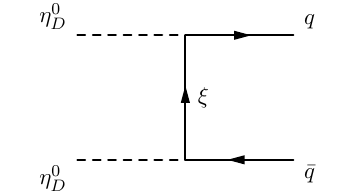} \\
	(A)\hspace{4.5cm}(B)\hspace{4.5cm}(C)
	\caption{Feynman diagrams for the (co)annihilation channels of the $Z_2$-odd singlet VLQ, $\xi$ that contributes to relic density. }
	\label{fig:feynRelic}
\end{figure}

\subsection{Vector-like Quark as Possible Solution}
The introduction of a vector-like quark in the particle-spectrum makes the phenomenology interesting by opening new (co)annihilations channels. The VLQ can potentially alter the relic density of the pure IDM like scenario significantly depending on the strength of the Yukawa-like coupling with the inert doublet and also the mass separation between the inert neutral Higgs (DM) and the VLQ. This additional VLQ modifies each of the three terms in eq.~\refeq{sigveffIDM} by mediating the processes in th $t$-channel which increases with $y_\xi$. And owing to the $Z_2$-odd nature, when the mass gap is small enough (co)annihilations of the VLQ affects the relic abundance of $\eta_D^0$. So, with the additional coannihilation channels the $\langle\sigma v\rangle_{eff}$ in eq.~\refeq{boltzmn} takes the form as 
\begin{eqnarray} 
	\label{sigveff}
	\langle\sigma v\rangle_{eff}&=& \langle\sigma v\rangle_{eff}^{\rm IDM+VLQ} \nonumber \\
	&& + \frac{2 g_0 g_\xi }{g^2_{\rm eff}}\langle\sigma v\rangle_{\eta_D^0 \xi} \Big(1+\Delta_{\xi}\Big)^{\frac{3}{2}} \, e^{-x \Delta_{\xi}}\nonumber \\
	&& + \frac{2 g_i g_\xi}{g^2_{\rm eff}}\langle\sigma v\rangle_{\eta_D^i\xi} \Big(1+\Delta_{\eta_D^i}\Big)^{\frac{3}{2}}\Big(1+\Delta_{\xi}\Big)^{\frac{3}{2}} e^{-x (\Delta_{\eta_D^i}+\Delta_{ \xi})}\nonumber \\
	&& + \frac{2 g^2_\xi}{g^2_{\rm eff}}\langle\sigma v\rangle_{\xi \xi} \Big(1+\Delta _{\xi}\Big)^3 \, e^{-2x \Delta_{\xi}}  \hspace{0.5cm}
\end{eqnarray}
	
\noindent where, $\langle\sigma v\rangle_{eff}^{\rm IDM+VLQ}$ is same as eq.~\refeq{sigveffIDM} except that contribution to all those terms will now include diagrams mediated by the VLQ $\xi$. The additional terms in the above equation correspond to the coannihilation contributions of the $Z_2$ odd scalars and VLQ whereas the last term represents the annihilation of the VLQ itself.
	
\begin{figure}[t!] 
	\centering
	\includegraphics[width=0.7\linewidth]{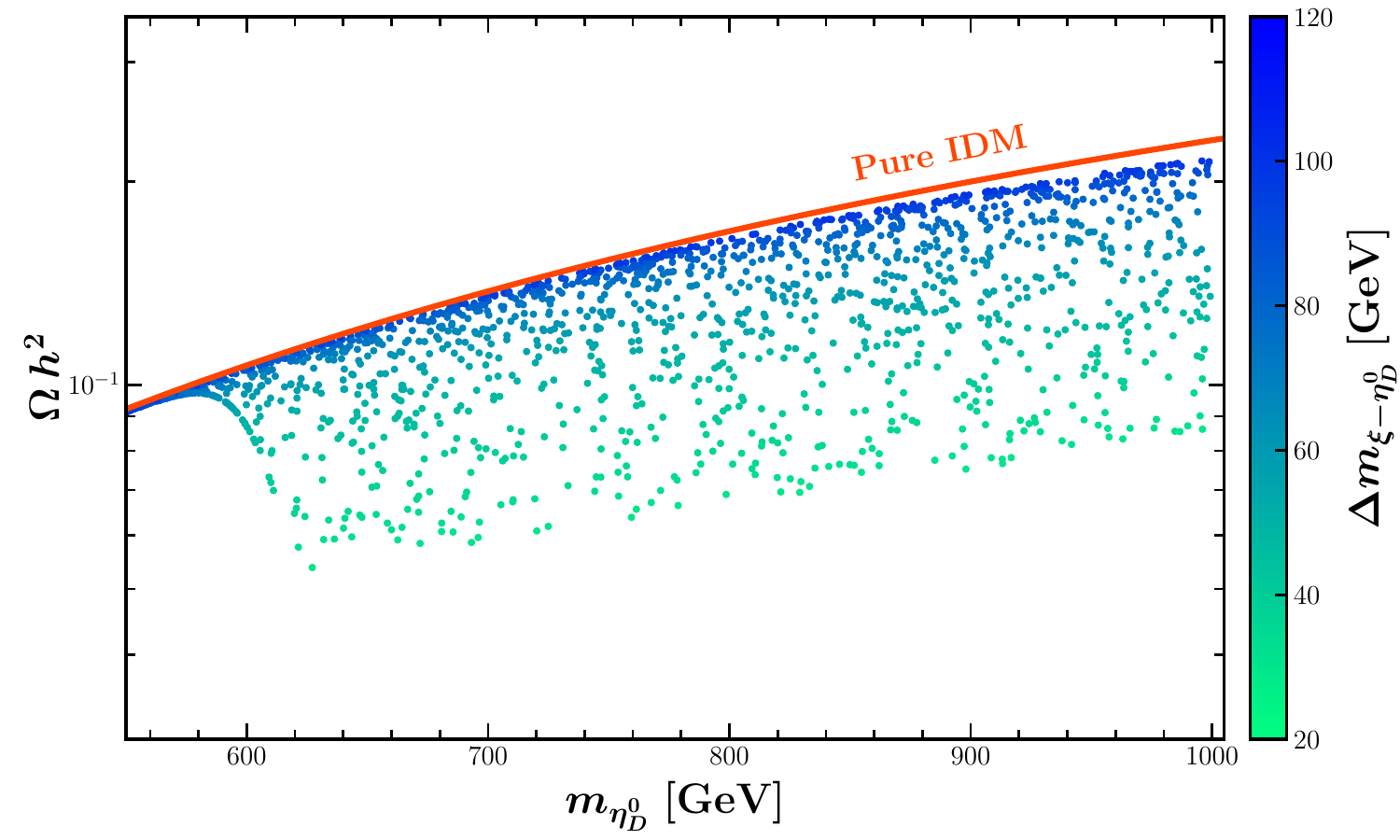}
	\caption{Variation of relic density with the DM mass in presence of VLQ compared with standard IDM case with a random scan over mass-splitting between DM and VLQ. Here $y_{\xi}=0.3$, $\lambda_L = 0.001$ and the mass gap between the charged and CP-odd scalar with the DM is kept fixed at 3 GeV and 5 GeV, respectively.}
	\label{fig:relicDmVLQ}
\end{figure}

Figure~\ref{fig:relicDmVLQ} depicts how the relic density varies with the mass-splitting between DM and the VLQ, $\Delta m_{\xi-\eta_D^0}$, across the DM mass range of our interest. We see that, for a fixed mass gap between the inert scalars, varying the mass separation between the DM and VLQ can lead to significant changes in the final relic density. 
From figure~\ref{fig:contIDM} we see, as the mass-splitting between the inert scalars is made  smaller the relic abundance gets larger, and eventually crosses the Planck-allowed value. To mitigate this we need to reduce the mass-gap of the VLQ and the DM.

\subsection{Contributions of VLQ} \label{subsec:vlq}	
\begin{figure}[t!]
	\centering
	\includegraphics[angle=0,width=7.3cm]{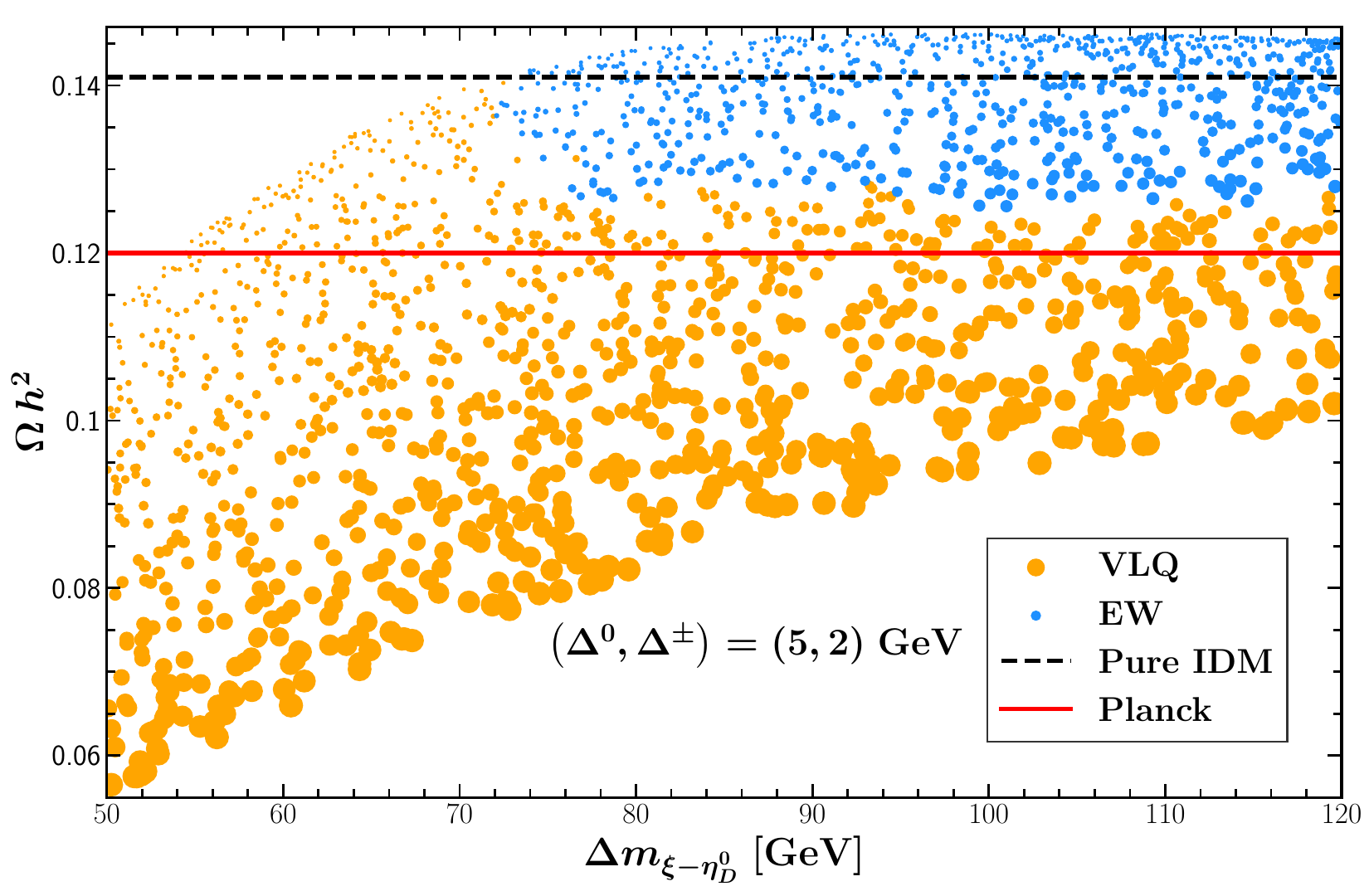} \hspace{0.1cm}
	\includegraphics[angle=0,width=7.3cm]{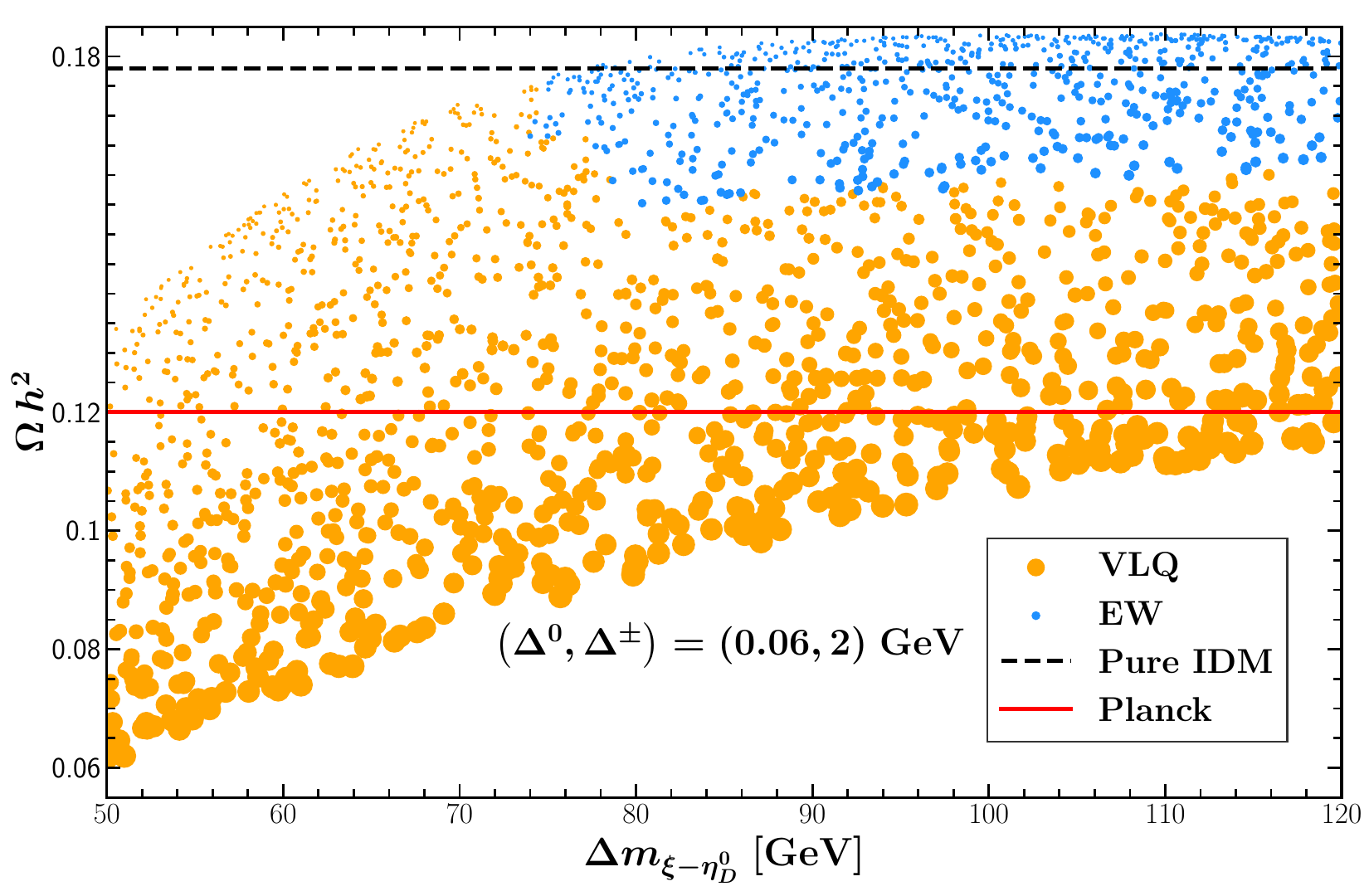}
	\caption{Scatter plot of DM relic density as a function of the mass-splitting between DM and VLQ ($\Delta m_{\eta_D^0\text{-}\xi}$) depicting the region that has at least $10\%$ contribution in the relic abundance from VLQ-induced processes. } 
	\label{fig:vlqEW}
\end{figure}

The contributions of VLQ to the DM relic can be estimated from the effective thermal averaged cross-section $\langle\sigma v\rangle_{eff}$ in eq.~\ref{sigveff}, and the contribution can be parameterized as a percentage value of $\langle\sigma v\rangle$ consisting the Feynman diagrams (figure~\ref{fig:feynRelic}) which include at least one VLQ particle, to the total $\langle\sigma v\rangle_{eff}$. In figure~\ref{fig:vlqEW} we show this contribution by dividing the event data set in two colors. The {\it orange} points denote the scan region where the VLQ contribution to $\langle\sigma v\rangle_{eff}$ is more than $10\%$ while the {\it blue} region corresponds to the region where it is less than $10\%$. Note that the point sizes differ in the scan as the size of these points denote the strength of the Yukawa coupling, which is varied in the range $0.1 < y_\xi < 0.9$. We find that for a fixed DM mass, the VLQ contribution to relic density is quite substantial for a large range of DM-VLQ mass separation and the Yukawa coupling.

To further analyze the role of VLQs in determination of the relic density we can separate the total parameter space in three distinct regions where the Feynman diagrams (figure~\ref{fig:feynRelic}) give contribution that have a particular mass hierarchy, as categorized in table~\ref{tab:diag_contri}.

\paragraph{Type-A --} One of the interesting features of having VLQ in the dark sector (DS) is the possibility of diluting the relic density of DM through VLQ self-annihilations through its strong interaction. Due to the expansion of the Universe, the temperature of the SM thermal bath decreases and at a certain temperature the (SM SM $\rightarrow$ DS DS) becomes energetically disfavored. Although at this point (DS DS $\rightarrow$ SM SM) remains active mainly because of the higher masses of DS particles provided the interaction responsible for (DS DS $\rightarrow$ SM SM) is faster than the Hubble expansion rate. As the DS particles remain in equilibrium with each other, if the number density of one species (${\rm DS}_A$) of DS decreases, the number density of rest of the species ${\rm DS}_i$ of DS will also decrease as they try to compensate the reduction of number density of the former species by converting into species A  (for example by the interaction ${\rm DS}_i \; {\rm SM}\,({\rm DS}_j) \rightarrow {\rm DS}_A \; {\rm SM}\,({\rm DS}_A)$). The exact amount of this conversion of DS species to species A depends on the assumption that the fraction of number density of each species to the total number density of DS remains constant $(r_i=n_{{\rm DS}_i}/n_{{\rm DS}}=n^{eq}_{{\rm DS}_i}/n^{eq}_{{\rm DS}})$ through out the evolution of Universe. Hence for a particular species A if this fraction $r_A$ decreases, it creates an out of equilibrium state in the DS, and rest of the particle number density will start to convert to the species A to restore the equilibrium. The process of conversion can happen until the interaction which is responsible for converting DS particles among each other can no longer overcome the Hubble expansion rate. 

\begin{figure}[t!]
	\centering
	\includegraphics[width=7.3cm]{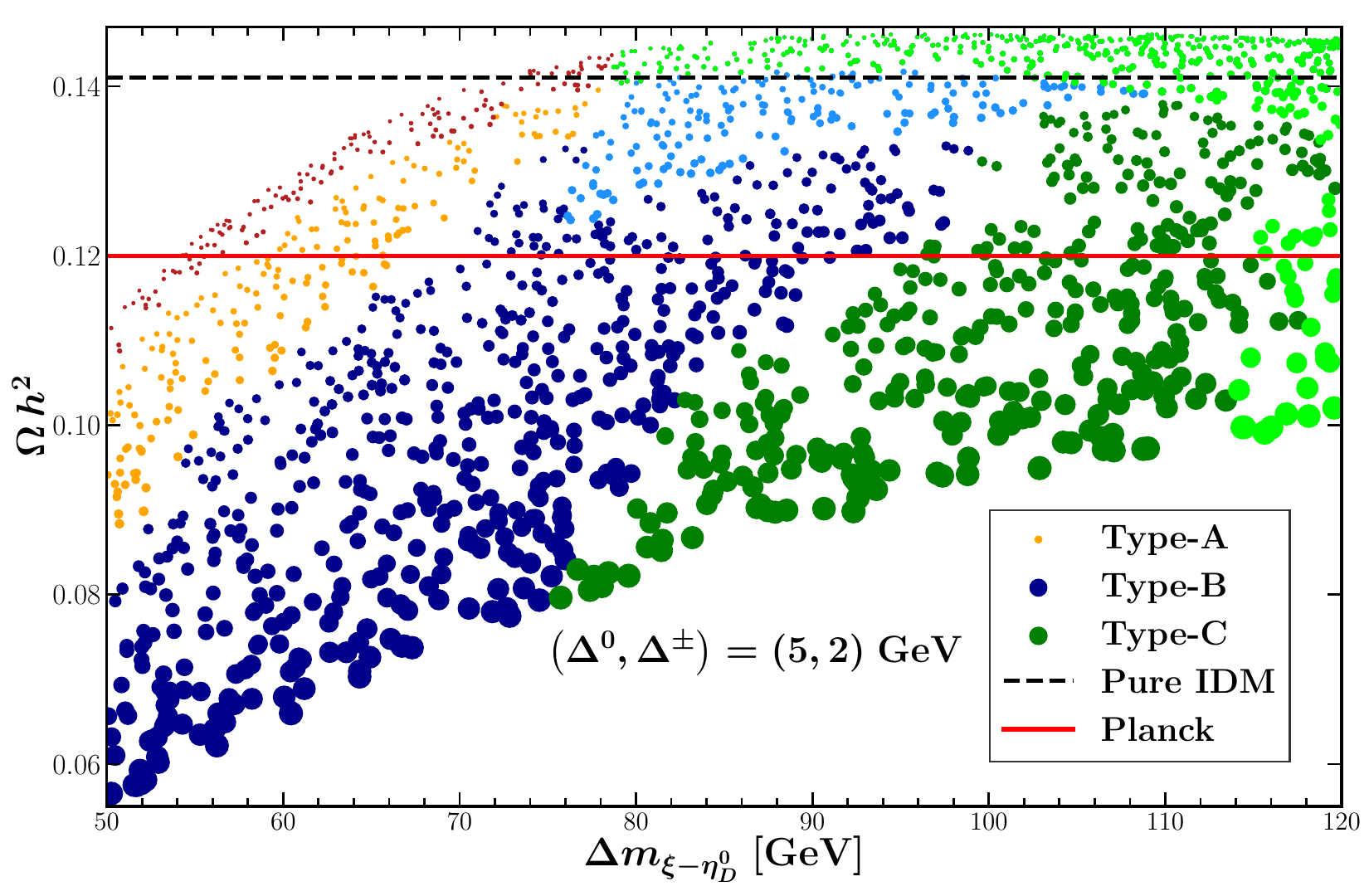} \hspace{0.1cm}
	\includegraphics[width=7.3cm]{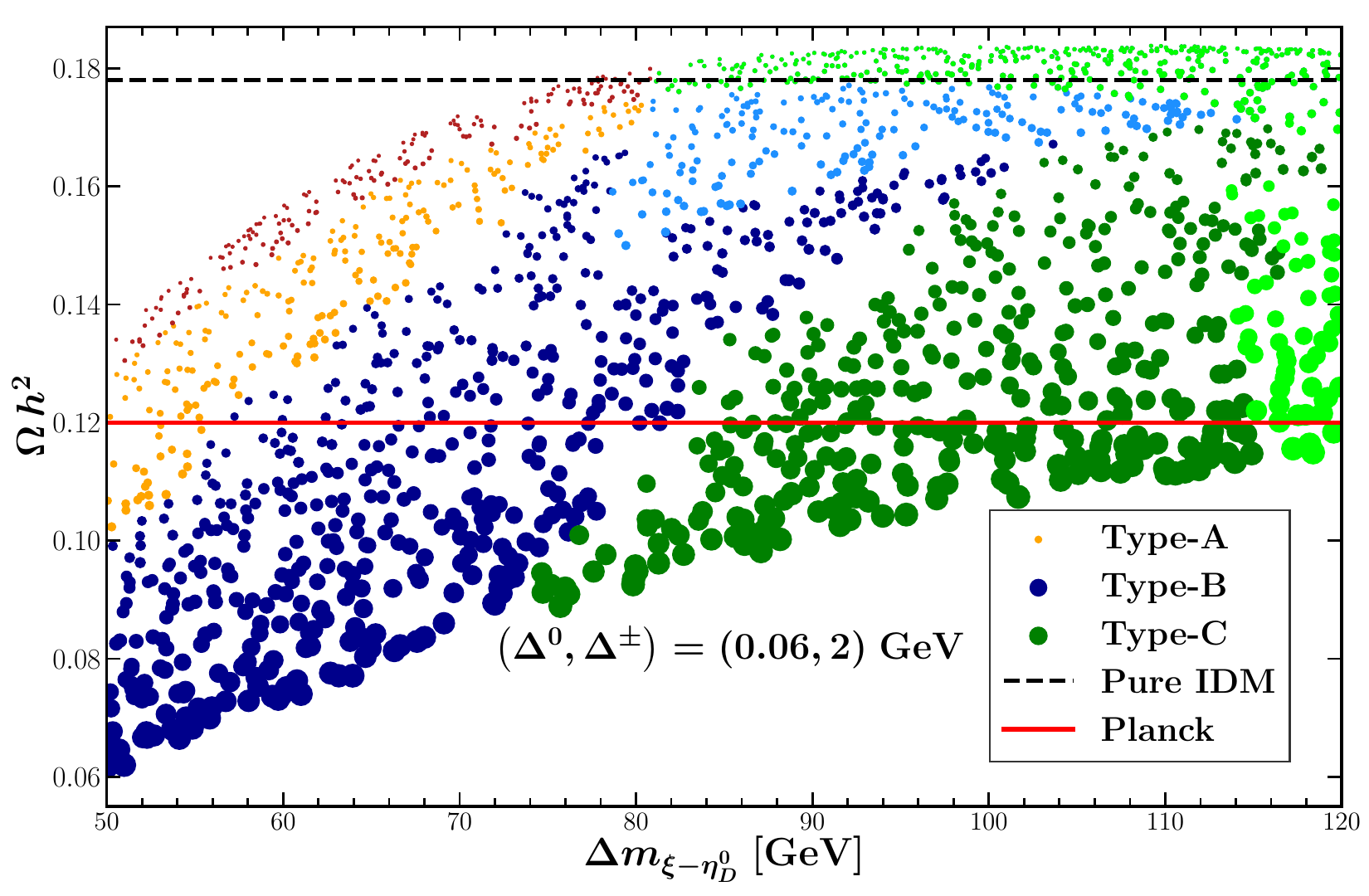}
	
	\caption{Scatter plot of relic density of DM against mass-splitting between DM and the VLQ ($\Delta m_{\xi-\eta_D^0}$). While the color of the points represent the Feynman diagram it gets the dominant contribution from, among all VLQ channels, the sizes of the points are scaled to the size of the Yukawa coupling $y_\xi$ which varies from 0.1 to 1.0.	} 
	\label{fig:3region}
\end{figure}

In our model because of the strong interaction of VLQ with SM, the process 
($\xi \; \xi \rightarrow$ SM SM) decreases its number density rapidly, hence one can expect that one DM ($\eta_D$) relic gets diluted because it is in equilibrium with $\xi$ as discussed above. In this region the diagram(A) contributes the most among all diagrams shown in figure~\ref{fig:feynRelic}. The role of this process is the reduction of the number of $Z_2$ odd species that could have contributed to the DM abundance. The large annihilation of the VLQ into SM quarks is a strong process and it gets an exponential suppression when the $\Delta_{\xi}$ is large, as seen in eq.~\ref{sigveff}. When the mass-separation between DM-VLQ is small and simultaneously the Yukawa coupling $y_\xi$ is chosen very small, the contribution of diagram(A) exceeds the combined contributions from diagram(B) and diagram(C). As the contribution from the latter two diagrams to the $\langle\sigma v\rangle_{eff}$ is proportional to $y_\xi^2$ and $y_\xi^4$, respectively they are relatively suppressed compared to diagram(A) that depends on the strong $SU(3)_c$ gauge coupling; hence, the relic density in this region has a strong dependence on DM mass and its mass-gap with the VLQ. We highlight this region in figure~\ref{fig:3region} with the red and orange points, where the red points denote contributions of diagram(A)  to be more than $90\%$ of the total VLQ induced subprocesses compared to diagram(B) and diagram(C), while the orange points highlight the region where its contribution is less than $90\%$, but is still greater than diagram(B) and diagram(C).    
	
\paragraph{Type-B --}
Here diagram(B) contributes the most as it spans over a parameter space where $y_\xi$ has moderate value ($\sim$0.6) and the DM-VLQ mass gap is also significant ($\sim65-90$ GeV). For this parameter region diagram(A) contributes the least in $\langle\sigma v\rangle_{eff}$ because of the extra  $e^{-2x \Delta_{\xi}}$ suppression compared to diagram(B). The contribution from diagram(C) is still suppressed in this region because of $y_\xi^4$ dependence. The light blue region marks the area where diagram(B) has more than $90\%$ contribution while the dark blue region represents the area where this contribution is still higher than others but less than $90\%$ of the total VLQ contribution.     
	
\paragraph{Type-C --}
In this case, diagram(C) dominates as it has no exponential suppression of DM-VLQ mass gap and varies only as ($\sim y^4_\xi/M^4_\xi$). Hence at large VLQ mass and $y_\xi$ coupling the contribution of this diagram falls slower than diagram(A) and diagram(B) because of no exponential suppression acting on it. We can also see that for a not so large value of $y_\xi$ there are few green points over blue region. This is due to  the VLQ particles decoupling from the thermal bath and their relic finally decays to DM after it has frozen. This boosts the DM relic density even above the value when pure IDM model is considered. We also find that below the light green point which denotes the region where it contributes more than $90\%$, and the dark green region  with less than $90\%$ contributions, consist of parameter space where diagram(C) gives the leading contribution.

\begin{table}[t!]
	\centering
	\resizebox{\linewidth}{!}{
		\begin{tabular}{|c|c|c|}
			\hline
			{\boldmath $y_\xi$} & {\boldmath $\Delta m_{\xi-\eta_D^0}$} & \textbf{Contributions from VLQ diagrams} \\ \hline
			Small ($<0.45$) & Small ($<70$ GeV) & Type-A $>$ Type-B $>$ Type-C \\ \hline
			Intermediate ($0.45-0.63$) & Intermediate ($70-90$ GeV) & Type-B $>$ Type-A $>$ Type-C \\ \hline
			Large ($>0.63$) & Large ($>90$ GeV) & Type-C $>$ Type-B $>$ Type-A \\ \hline
		\end{tabular}
	}
	\caption{Different ranges of the VLQ-related parameter and the order of the corresponding contributions from the three types of Feynman diagrams as in Figure~\ref{fig:3region} for $M_{\rm DM}=700 \text{ GeV},~ \Delta^0=5 \text{ GeV},~\Delta^\pm=2 \text{ GeV} $ with $\lambda_L=0$ while satisfying DM relic density.} 
	\label{tab:diag_contri}
\end{table}

In Table~\ref{tab:tab3bp}, we present few benchmark points which illustrate the advantages of having a VLQ in the IDM model compared to pure IDM scenario. The VLQ aids the DM annihilation cross-section, and thus removes the reliance on $\lambda_L$ in decreasing the DM relic density in the over abundant region. In BP3 and BP4 we see for the chosen dark sector mass spectrum, to get the correct relic density in pure IDM case we require high value of $\lambda_L$, which further increases the direct detection cross-section, resulting it getting ruled out by the current DD bounds. The addition of VLQ achieves the correct DM abundance without depending on such high value of $\lambda_L$, effectively making the DD cross-section small and within experimentally observed limits.

\begin{table}[t!]
	\centering
	\resizebox{\linewidth}{!}{
		\begin{tabular}{|c|c|c|c|c|c|c|c|c|c|c|}
			\hline
			\multirow[m]{2}{0.5cm}{\centering\textbf{BP}} & \boldmath $m_{\eta_D^0}$ & \boldmath $\Delta^0$ & \boldmath $\Delta^\pm$ & \multirow[m]{2}{0.5cm}{\centering\boldmath $\lambda_L$} & \multicolumn{2}{c|}{\textbf{IDM}} & \boldmath $\Delta m_{\xi\text{-}\eta_D^0}$ & \multirow[m]{2}{0.5cm}{\centering\boldmath $y_\xi$} & \multicolumn{2}{c|}{\textbf{IDM+VLQ}}  \\ \cline{6-7}\cline{10-11}
			& \textbf{(GeV)} & \textbf{(GeV)} & \textbf{(GeV)} && {\boldmath $\Omega h^2$} & {\centering\boldmath $\sigma_{\eta_D^0\text{-}N}^{SI}\,(\textbf{cm}^2$)}& \textbf{(GeV)} &&{\centering\boldmath $\Omega h^2$} & {\centering\boldmath $\sigma_{\eta_D^0\text{-}N}^{SI}\,(\textbf{cm}^2$)} \\ \hline
			
			BP-1 & 1000 & 0.5 & 1.0 & 0.001 & 0.365 & $3.59\times 10^{-49}$ & 50 & 0.60 & 0.119 & $3.56\times 10^{-47}$ \\ \hline
			BP-2 & 1000 & 0.5 & 1.0 & 0.059 & 0.305 & $3.18\times 10^{-46}$ & 50 & 0.55 & 0.121 & $2.40\times 10^{-47}$ \\ \hline
			BP-3 & 1000 & 1.0 & 1.0 & 0.238\,(0.03) & 0.114 & $9.96\times 10^{-46}$ & 35.5 & 0.01 & 0.119 & $1.64\times 10^{-47}$ \\ \hline
			BP-4 & 800 & 1.0 & 1.0 & 0.155\,(0.03) & 0.114 & $ 7.033\times 10^{-46}$ & 42.0 & 0.01 & 0.12 & $1.65\times 10^{-47}$ \\ \hline

		\end{tabular}
	}
	\caption{Summary table depicting some benchmark points where the constraints from direct detection and relic density are not satisfied simultaneously in the pure IDM; however, the singlet VLQ in the model helps achieve it.}
	\label{tab:tab3bp}
\end{table}

\subsection{Collider Outlook}

\begin{figure}[t!]
	\centering
	\includegraphics[width=0.7\linewidth]{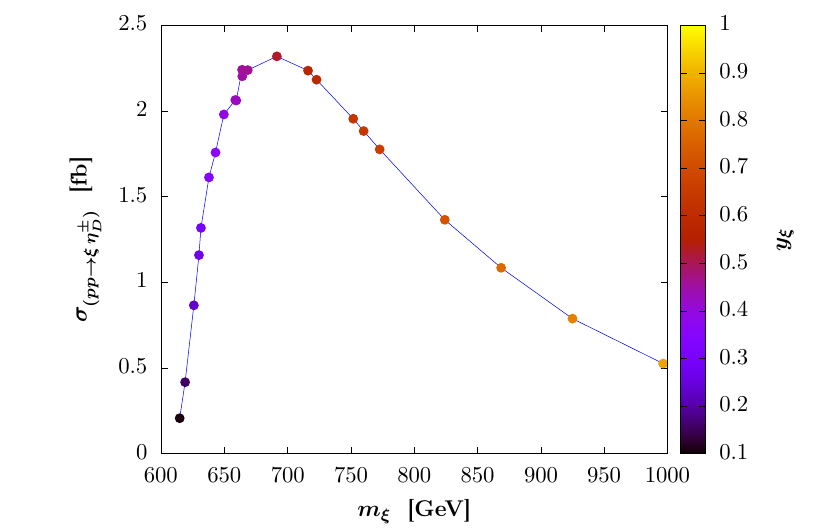}
	\caption{Showing the associated production cross-section of $p p \to \bar{\xi} \eta_D^+ + \xi  \eta_D^-$ at the 14 TeV LHC as a function of the VLQ mass.  The IDM sector is set at a fixed mass gap $(\Delta^0, \Delta^\pm) = (0.5,1)$ GeV while the VLQ-DM mass gap is fixed at 60 GeV. The colored points on the curve represent the strength of the interaction strength ($y_\xi$) for a given $\xi$ mass that gives the correct DM relic abundance while also satisfying the indirect and direct search constraints.}
	\label{fig:associated_prod}
\end{figure}

Finally, we examine in this section how the introduction of the VLQ in the IDM can affect the search strategies for the BSM particles in collider experiments. In hadron colliders, like the LHC, the VLQs 
can be produced singly and in pairs. As the VLQ's are odd under the discrete symmetry $Z_2$, their 
single production will always have an associated IDM scalar. In our case, as we have focused on the 
top-like VLQ, the only relevant channel of single production would be $p p \to \bar{\xi} \eta_D^+ + \xi  \eta_D^-$. This process is expected to proceed mainly via the gluon-bottom fusion as the remaining subprocesses will be CKM suppressed. As the production cross-section will depend on the coupling strength $y_\xi$, its choice will be crucial in determining the event rate for this process. 
In figure \ref{fig:associated_prod} we show the associated production cross-section changing as we vary the VLQ mass and $y_\xi$. We used the \texttt{UFO}~\cite{Darme:2023jdn} library obtained using \texttt{SARAH} in \texttt{MadGraph5}~\cite{Alwall:2014hca} to obtain the cross-sections. The plot highlights some interesting details, where the shown curve represents the associated production cross-section for values of $\xi$ mass and coupling $y_\xi$, only when the DM has the correct relic abundance and is within the $3\sigma$ error limits.  Here the VLQ-DM mass gap is kept fixed at $\Delta_\xi =60$ GeV. As the DM relic abundance also depends on the $\Delta_\xi$ we will get multiple curves for different values of $\Delta_\xi$ in figure \ref{fig:associated_prod}. We have shown only one case as a representative curve.  As we increase the DM mass its relic abundance also increases because the annihilation processes in figure \ref{fig:feynRelic} get suppressed because of higher VLQ mass. Hence to increase the cross-section  of the annihilation channels  we require larger values of $y_\xi$ to satisfy the DM relic abundance. 
On the other hand, the cross-section which depends on $y_\xi$  also becomes larger. One must however note that at very large $\xi$ mass, the charged-scalar is also as heavy and the phase space suppression kicks in. Therefore, after a given $m_\xi$ the enhancement in the production cross-section due to larger $y_\xi$ is counterbalanced and the cross-section is found to start decreasing.

\begin{figure}[t!]
	\centering
	\includegraphics[width=0.6\linewidth]{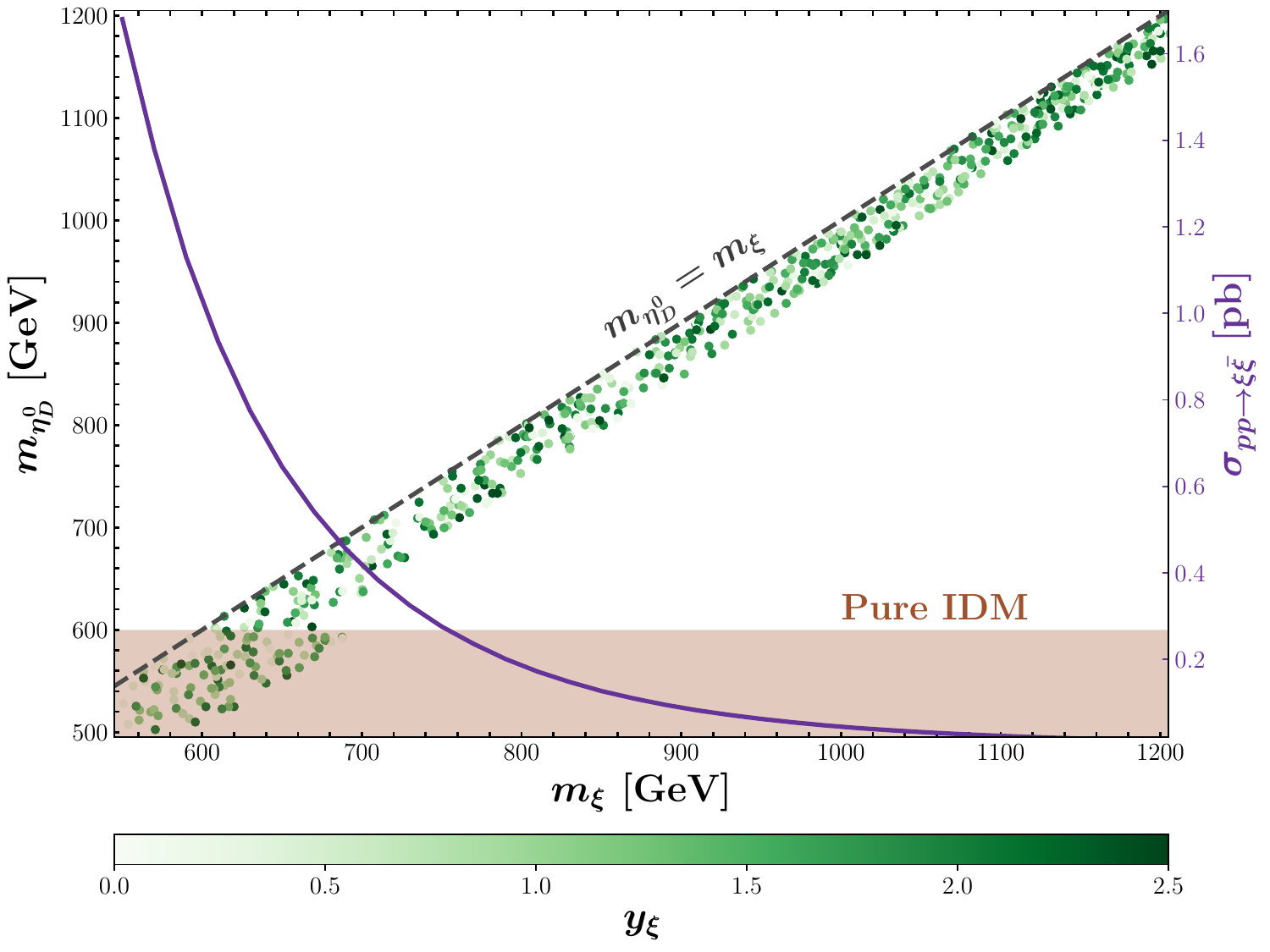}
	\caption{Scatter plot of the relic allowed parameter space as function of DM mass and VLQ mass for a random scan over $y_\xi$ for $(\Delta^0, \Delta^\pm) = (5,3)$ GeV. The shaded region is for the pure IDM scenario where the VLQ does not affect the DM phenomenology. The violet-colored line plot gives the variation of the pair-production cross-section of VLQ at 14 TeV LHC as a function the VLQ mass.}
	\label{fig:collider}
\end{figure}

The other production mode is the dominant pair-production via strong interactions. At lepton colliders, the VLQ can be pair produced due to its non-zero coupling with the photon and $Z$ boson. These VLQs will subsequently decay to produce a SM quark and one of the inert scalars, resulting in missing energy signatures in the detectors. As we have considered the VLQ that couples to the third generation SM quark doublet, we will either produce top quarks or bottom quarks as its decay by-products. This will lead to several possibilities for the new physics signal at the LHC, and in the presence of VLQ we can enhance the inert scalar productions substantially compared to the pure IDM scenario. 
The pair production cross-section of the VLQ at 14 TeV LHC is shown in figure~\ref{fig:collider} as a function of its mass with the violet line. The production cross-section mainly depends on the VLQ mass.  As already discussed in section~\ref{subsec:vlq}, for sufficiently high mass-gap between the VLQ and DM, the role of (co)annihilation of VLQ, in $\langle\sigma v\rangle_{eff}$ is exponentially suppressed. Also, depending upon the mass-splitting between the inert scalars we need to tune the mass gap between the $\xi$ and $\eta_D^0$. In figure~\ref{fig:collider} we show this behavior with a scatter plot in the $m_\xi-m_{\eta_D^0}$ plane where the green points represent the region where the relic density is satisfied or underabundant by the combination of inert scalar and VLQ. The  horizontal shaded region at the bottom represents the pure IDM case, where the VLQ contribution is ignored. The darker points represent higher values of $y_\xi$ as represented by the colorbar at the bottom. 

Thus the VLQ can play a significant role in DM phenomenology and has robust collider signals. In the current work we have highlighted that the VLQ production at LHC, if dictated by DM favoured parameter space will lead to a compressed spectrum with the VLQ decaying to third generation quarks and the inert scalars. As the decay proceeds through a Yukawa coupling, the decay could be prompt or even make the VLQ long-lived. We leave a more detailed collider analysis of the model for future work.

\section{Conclusions}\label{sec:concl}

The IDM is a popular model for DM study because of its simple nature yet rich phenomenology. Due to its $Z_2$-odd nature it offers a plausible DM candidate. The IDM has three distinct regions in the DM mass range, namely the low mass region where the relic density is large due to interaction with SM Higgs; the intermediate region with small relic density due to high annihilation rate into vector bosons, and the high mass region that again increases due to a cancellation of the contributions between the four-point interaction and a $t$-channel mediated subprocess, eventually making it overabundant. Introducing a VLQ which is also odd under $Z_2$ gives an additional leverage to the dark sector to expand the allowed parameter space of the IDM candidate for DM. We find that due to the VLQ's Yukawa interaction with the inert doublet and SM quarks and its strong interaction with the gluon, it can influence the DM phenomenology significantly.

In this work, we show how an additional VLQ in the model helps to improve the relic density scenario of the erstwhile IDM in mass regions disfavored in the pure IDM scenario, while evading direct detection constraints when the new scalar masses are nearly degenerate.  As the interaction of the DM with the SM Higgs ($\lambda_L$) contributes to the relic abundance as well as to the DD cross-section,  $\lambda_L$ has to be very small to evade the DD constraints. Also the mass gap between DM and other inert scalars can be made very small for improved co-annihilations but certain interference effects become dominant to negate this. Thus, in the high mass range of DM ($\gtrsim 550$ GeV)
both of these scenarios make the relic abundance go beyond the Planck-observed value. The addition of the $Z_2$-odd VLQ helps in improving the situation by opening new (co)annihilation channels of the DM as discussed in section~\ref{subsec:vlq}. A crucial test for the existence of these VLQ's can come from collider searches. We determine that the bounds on a $Z_2$ odd third generation VLQ are in the ballpark of the mass of the DM in the IDM scenario where the relic starts becoming overabundant. Therefore an immediate benefit of having the exotic quark is that it improves the DM scenario and are within the reach of the LHC. The VLQ also helps in boosting the production of these inert scalars either through its decay or via associated production. A detailed analysis might reveal the origin of the VLQs and the scalar DM. In summary, the inert doublet model extended by a vector-like quark  makes the compressed IDM dark matter option viable at higher mass region, producing allowed relic abundance while relaxing the direct detection constraints. The scenario presented in this work also highlights the potentially interesting collider signatures that could be observed at present and future colliders.

\section*{Acknowledgments}
SD, SK and SKR would like to acknowledge the support from Department of Atomic Energy (DAE), Government of India for the Regional Centre for Accelerator based Particle Physics (RECAPP), Harish-Chandra Research Institute.
\bibliographystyle{JHEP}
\bibliography{Reference}%

\end{document}